\documentclass{PoS}

\makeatletter

\def\lsim{\compoundrel<\over\sim}
\def\compoundrel#1\over#2{\mathpalette\compoundreL{{#1}\over{#2}}}
\def\compoundreL#1#2{\compoundREL#1#2}
\def\compoundREL#1#2\over#3{\mathrel
	{\vcenter{\hbox{$\m@th\buildrel{#1#2}\over{#1#3}$}}}}
\makeatother

\title{%Contribution title
Hyperon-nucleon potentials from lattice QCD
}

\ShortTitle{%Short Title for header
Hyperon-nucleon potentials from lattice QCD
}

\author{\speaker{%First Author
Hidekatsu Nemura
}%\thanks{A footnote may follow.}
\\
	Advanced Meson Science Laboratory,
	Nishina Center for Accelerator-Based Science, RIKEN,% \\
	Wako, %Saitama,
	351-0198, Japan \\
        E-mail: \email{%author@email
	nemura@riken.jp
}}

\author{%Another Author\\
	Noriyoshi Ishii\\
	Center for Computational Sciences,
	University of Tsukuba, %\\
	Tsukuba 305-8571, Japan \\
	E-mail: \email{ishii@rarfaxp.riken.jp}
	}

\author{%Another Author\\
	Sinya Aoki\\
	Graduate School of Pure and Applied Sciences,
	University of Tsukuba, %\\
	Tsukuba 305-8571, Japan \\
	Riken BNL Research Center, Brookhaven National Laboratory, %\\
	Upton, New York 11973, USA \\
	E-mail: \email{saoki@het.ph.tsukuba.ac.jp}
	}

\author{%Another Author\\
	Tetsuo Hatsuda\\
	Department of Physics, University of Tokyo, %\\
	Tokyo 113-0033, Japan \\
	E-mail: \email{hatsuda@phys.s.u-tokyo.ac.jp}
	}

\abstract{%..........................\
	  We calculate $p\Xi^0$ potentials from the equal-time
	  Bethe-Salpeter amplitude measured in the
	  quenched QCD simulation with the spatial lattice volume,
	  (4.4 fm)$^3$.
	  The standard Wilson gauge action with the gauge coupling
	  $\beta=5.7$ on $32^4$ lattice together with the standard
	  Wilson quark action are used. 
	  The hopping parameter $\kappa_{ud}=0.1678$ is chosen for 
	  $u$ and $d$ quarks,
	  which corresponds to $m_{\pi}\simeq 0.37$ GeV.
	  The physical strange quark mass is used
	  by taking the parameter $\kappa_s=0.1643$ which is 
	  deduced from the physical $K$ meson mass.
	  The lattice spacing $a=0.1420$ fm is determined by 
	  the physical $\rho$ meson mass.
	  We find that
	  the $p\Xi^0$ potential 
	  has strong spin dependence.
	  Strong repulsive core is found in $^1S_0$ channel 
	  while the effective central potential in the $^3S_1$ channel 
	  has relatively weak repulsive core.
	  The potentials also have weak attractive parts 
	  in the medium to long distance region 
	  ($0.6$ fm $\lsim r \lsim 1.2$ fm) 
	  in both of the $^1S_0$ and $^3S_1$ channels. 
	  }

\FullConference{The XXV International Symposium on Lattice Field Theory\\
		 July 30-4 August 2007\\
		 Regensburg, Germany}

\begin{document}

\section{Introduction}

Study of hypernuclei is one of the frontiers 
in nuclear physics. 
The strange degree of freedom gives a new dimension to 
the description of nuclear structure. 
A hyperon (or a strange quark) embedded in nuclei 
plays a characteristic 
role as an ``impurity'' or a probe in the many body system\cite{Tamura}. 

Modern nucleon-nucleon ($NN$) potentials give 
a successful description of the $NN$ scattering data
and have been used to make precise calculations of 
light nuclei\cite{Nogga,Pieper}. 
In contrast, 
hyperon-nucleon ($YN$) and hyperon-hyperon ($YY$) interactions 
have large uncertainties, because the scattering experiments 
are either difficult or impossible due to the short life-time of 
hyperons. 
The full phase shift analysis at the same level of the $NN$ case
is not available yet. 
Although a lot of theoretical models which describe $YN$ 
interaction 
together with 
the $NN$ interaction have been 
published, the different model predicts different phase shifts 
and scattering parameters, e.g., of the $\Lambda N$ potential 
in spite of the nice description of the $NN$ 
sector\cite{ESC04,NSC97,NSC89,FSS,fss2,GSOBEP
}.

Recently, a lattice QCD study of the $NN$ potential has been 
performed\cite{Ishii}.  
This approach may lead to a new 
paradigm to study the 
$YN$ and $YY$ interactions too, 
since the lattice QCD is an {\it ab initio} method of 
treating the fundamental theory of strong interaction. 
(See also the conventional approach to the $YN$ phase shifts using the
L\"{u}scher's finite volume formula\cite{Beane}.)

The purpose of the present report
is to explore the $YN$ and $YY$ potentials 
from the lattice QCD simulation 
 on the basis of the methodology developed in
 Refs.~\cite{Ishii,Aoki}.
The extension from the $NN$ potential 
to $YN$ and $YY$ potentials is relatively 
straightforward. 
In the case of the $NN$ potential, 
there are only two representations in the isospin channel, i.e., 
${\bf 2}\otimes{\bf 2}={\bf 3}\oplus{\bf 1}$ in $SU(2)$,
which correspond to isovector and isoscalar channels. 
Including the strange degrees of freedom extends the arithmetic 
into flavor $SU(3)$, 
$
{\bf 8}\otimes{\bf 8}=
 {\bf 27}\oplus\overline{\bf 10}\oplus{\bf 1}\oplus{\bf 8}\oplus
 {\bf 10}\oplus{\bf 8}. 
$
Here the isovector (isoscalar) channel of the $NN$ sector is 
assigned to be a subset in the ${\bf 27}$-plet 
($\overline{\bf 10}$-plet) representation. 
The potentials for newly arising channels 
are hardly determined from the real experiment so far. 
The lattice QCD simulation 
with physical strange quark mass 
provides 
new numerical 
``data'' for 
these strange 
channels.  

In this report, 
we focus on the $N\Xi$ potential 
in the isovector ($I=1$) channel 
as a first step. 
We keep away, at present, from the isoscalar ($I=0$) channel 
of the $N\Xi$ potential, 
since it is not the lowest state 
of the isoscalar $^1S_0$ channel and 
$\Lambda\Lambda$ strong decay mode may open below the $N\Xi$. 
There are almost no experimental information on the 
$N\Xi$ interaction, 
although 
a few experimental data\cite{Nakazawa,Fukuda,Khaustov} 
suggest that the 
$\Xi$-nucleus potential would be weakly attractive. 
Moreover, 
the $\Xi$-nucleus interaction will be studied 
as a day one experiment in the near future 
at J-PARC\cite{JPARC} through $(K^-,K^+)$ reaction on 
the nuclear target such as $^{12}$C.

\section{Formulation}

As we mentioned, the methodology to obtain the potential is
along the lines of Refs.~\cite{Aoki} and \cite{Ishii}.  
The latter 
describes, successfully for the first time, 
the $NN$ potentials from lattice QCD simulation. 
We start from an effective Schr\"{o}dinger equation
for $N\Xi$ system at low energies:
\begin{equation}
 -{1\over 2\mu}\nabla^2 \phi(\vec{r}) +
  \int d^3r^\prime U(\vec{r},\vec{r}^\prime)
  \phi(\vec{r}^\prime) =
  E \phi(\vec{r}), 
\end{equation}
where $\mu=m_{N}m_{\Xi}/(m_{N}+m_{\Xi})$ and 
$E$ are the reduced mass of the $N\Xi$ system and 
the nonrelativistic energy in the center-of-mass frame, respectively. 
The nonlocal potential can be represented by the
derivative expansion as
\begin{equation}
U(\vec{r},\vec{r}^\prime)=
 V_{N\Xi}(\vec{r},\nabla)\delta(\vec{r}-\vec{r}^\prime).
\end{equation}
The general expression of the potential $V_{N\Xi}$ 
would be 
\begin{eqnarray}
 V_{N\Xi} &=&
  V_0(r)
  +V_\sigma(r)(\vec{\sigma}_N\cdot\vec{\sigma}_\Xi)
  +V_\tau(r)(\vec{\tau}_N\cdot\vec{\tau}_\Xi)
  +V_{\sigma\tau}(r)
   (\vec{\sigma}_N\cdot\vec{\sigma}_\Xi)
   (\vec{\tau}_N\cdot\vec{\tau}_\Xi)
   \nonumber \\
 &&
  +V_T(r)S_{12}
  +V_{T\tau}(r)S_{12}(\vec{\tau}_N\cdot\vec{\tau}_\Xi)
  +V_{LS}(r)(\vec{L}\cdot\vec{S}_+)
  +V_{LS\tau}(r)(\vec{L}\cdot\vec{S}_+)(\vec{\tau}_N\cdot\vec{\tau}_\Xi)
  \nonumber \\
 &&
  +V_{ALS}(r)(\vec{L}\cdot\vec{S}_-)
  +V_{ALS\tau}(r)(\vec{L}\cdot\vec{S}_-)(\vec{\tau}_N\cdot\vec{\tau}_\Xi)
  +{O}(\nabla^2). 
\end{eqnarray}
Here
$S_{12}=3(\vec{\sigma}_1\cdot\vec{n})(\vec{\sigma}_2\cdot\vec{n})-\vec{\sigma}_1\cdot\vec{\sigma}_2$
is the tensor operator with $\vec{n}=\vec{r}/|\vec{r}|$,
$\vec{S}_{\pm}=(\vec{\sigma}_1 \pm \vec{\sigma}_2)/2$ 
symmetric ($+$) and antisymmetric ($-$) spin operators,
$\vec{L}=-i\vec{r}\times\vec{\nabla}$ the relative 
angular momentum operator, 
and 
$\vec{\tau}_N$ ($\vec{\tau}_{\Xi}$) is isospin operator 
for $N$ ($\Xi$). 
We note that the antisymmetric spin-orbit forces 
($V_{ALS}$ and $V_{ALS\tau}$) 
newly come up because the constituents ($N$ and $\Xi$) are 
not identical.

According to the above expansion of the potential, 
the wave function should be classified by the total isospin $I$, 
the total angular momentum and parity $J^\pi$ with 
$\vec{J}=\vec{L}+\vec{S}_+$. 
Particular spin (isospin) projection can be made in terms of 
$\vec{\sigma}_N\cdot\vec{\sigma}_\Xi$
($\vec{\tau}_N\cdot\vec{\tau}_\Xi$), e.g., 
for the isospin projection we have 
$P^{(I=0)}=(1-\vec{\tau}_N\cdot\vec{\tau}_\Xi)/4$ 
and 
$P^{(I=1)}=(3+\vec{\tau}_N\cdot\vec{\tau}_\Xi)/4$. 
In this work, we focus only the isospin $I=1$,
$S$-wave component of the wave function, 
so as to obtain the (effective) central potential through 
\begin{equation}
 V_{\rm central}(r) = E +
  {1\over 2\mu}{\vec{\nabla}^2\phi(r)\over \phi(r)}. 
\end{equation}

The $S$-wave wave function 
is measured from 
the equal-time Bethe-Salpeter (BS) amplitude
$\phi(\vec{r};k)$ 
as 
\begin{eqnarray}
 \phi(\vec{r};k) &=&
  {1\over 24} \sum_{{\cal R}\in O} {1\over L^3} \sum_{\vec{x}}
  P^\sigma_{\alpha\beta} 
  \left\langle 0
   \left|
    p_\alpha({\cal R}[\vec{r}]+\vec{x})
    \Xi^0_\beta(\vec{x})
   \right| p \Xi^0 ; k
  \right\rangle,
  \\
  p_\alpha(x) &=&
  \varepsilon_{abc} \left(
		     u_a(x) C \gamma_5 d_b(x)
		    \right)
  u_{c\alpha}(x),
  \\
  \Xi^0_\beta(y) &=&
  \varepsilon_{abc} \left(
		     u_a(y) C \gamma_5 s_b(y)
		    \right)
  s_{c\beta}(y),
\end{eqnarray}
where $\alpha$ and $\beta$ denote the Dirac indices, 
$a$, $b$ and $c$ the color indices, and 
$C=\gamma_4\gamma_2$ the charge conjugation matrix.
The summation over ${\cal R}\in O$ is taken for cubic transformation 
group to project onto the $S$-wave, and
the summation over $\vec{x}$ for zero total spatial momentum. 
$p_\alpha(x)$ and $\Xi^0_\beta(y)$ are the local field operators
for the proton and $\Xi^0$. 
We take the upper components of the Dirac indices $\alpha$ and $\beta$
to construct the spin singlet (triplet) channel by 
$P^\sigma_{\alpha\beta}=(\sigma_2)_{\alpha\beta}$
($P^\sigma_{\alpha\beta}=(\sigma_1)_{\alpha\beta}$).

The $\phi(\vec{r})$ with $\vec{r}=\vec{x}-\vec{y}$
is understood as the probability amplitude to find 
``nucleon-like'' three quarks located at point $\vec{x}$ and
``$\Xi$-like'' three quarks located at point $\vec{y}$.
The 
$\phi(\vec{r})$ 
includes not only the
elastic amplitude $N\Xi\rightarrow N\Xi$ but also the
inelastic amplitudes such as $N\Xi\rightarrow \pi N\Xi$ and 
$N\Xi \rightarrow \Lambda\Sigma$, and so on. 
Note that, at low energies below the thresholds, however, 
the asymptotic behavior of $\phi(\vec{r})$ 
is not affected by the inelastic contributions,
since they decrease exponentially 
in the asymptotic region. 
{
(In the present calculation with the isospin $I=1$,
the $\Lambda\Lambda$ channel is closed
due to the isospin conservation.)
}
On the other hand, 
$\phi(\vec{r})$ 
and hence the potential 
 may depend on the 
interpolating fields 
in the interaction region. 
Further study on this issue is 
found in Ref.\cite{IshiiFull} for 
the $NN$ potential.

In the actual simulations, the BS amplitude is obtained through the 
four-point 
correlator,
\begin{eqnarray}
 F_{p\Xi^0}(\vec{x}, \vec{y}, t; t_0) &=&
  \left\langle 0
   \left|
    p_\alpha(\vec{x},t)
    \Xi^0_\beta(\vec{y},t)
    \overline{\cal J}_{p \Xi^0}(t_0)
   \right| 0 
  \right\rangle
  \\
 &=&
  \sum_n A_n
  \left\langle 0
   \left|
    p_\alpha(\vec{x})
    \Xi^0_\beta(\vec{y})
    \right| n
  \right\rangle
  {\rm e}^{-E_n(t-t_0)}.
\end{eqnarray}
Here $\overline{\cal J}_{p\Xi^0}(t_0)$ is a source term located at
$t=t_0$. 
We utilize the wall source for $\overline{\cal J}_{p\Xi^0}$ in this work 
in order to enhance the lowest scattering state of the $p\Xi^0$ system. 
$E_n$ is the energy of the $p\Xi^0$ state, $|n\rangle$,
and $A_n(t_0)=\langle n|\overline{\cal J}_{p\Xi^0}(t_0)|0\rangle$.

We also calculate the two-point 
correlator, 
$C(t;t_0)=\sum_{\vec{x}}\langle 0|{\cal B}_\alpha(\vec{x},t)
\overline{\cal B}_\alpha(\vec{x}^\prime,t_0)|0\rangle$, 
for the octet baryons (${\cal B}=N,\Xi,\Lambda,\Sigma$), 
in order to check whether various two baryon 
(${\Lambda\Lambda}$,${N\Xi}$,${\Lambda\Sigma}$ and ${\Sigma\Sigma}$) 
thresholds are reproduced in the correct order. 
The interpolating fields for $\Lambda$ and $\Sigma^+$, 
employed in this work, 
are given by
\begin{equation}
 \Lambda_\alpha(x)=
  \varepsilon_{abc} \left\{
		     \left(
		      d_a(x) C \gamma_5 s_b(x)
		     \right) u_{c\alpha}(x)
		     + \left(
			s_a(x) C \gamma_5 u_b(x)
		       \right) d_{c\alpha}(x)
		     - 2 \left(
			  u_a(x) C \gamma_5 d_b(x)
			 \right) s_{c\alpha}(x)
		    \right\},
\end{equation}
\begin{equation}
 \Sigma^+_\beta(y)=
  - \varepsilon_{abc} \left(
		     u_a(y) C \gamma_5 s_b(y)
		    \right)
  u_{c\beta}(y).
\end{equation}

\section{Numerical calculation}

We use the standard Wilson gauge action at the gauge coupling
$\beta=5.7$ on the $32^3\times 32$ lattice together with the
standard Wilson quark action. 
See Ref.\cite{IshiiFull} for details. 
The hopping parameter of $\kappa_{ud}=0.1678$ is chosen for the 
$u$ and $d$ quarks, 
which corresponds to $m_\pi\simeq 0.37$~GeV, $m_\rho\simeq0.81$~GeV, 
and $m_N\simeq 1.16$~GeV. 
In order to determine the parameter for the strange quark mass ($\kappa_s$), 
we first measure the 
correlators of pseudo scalar and vector 
mesons by using the interpolating fields given by
\begin{eqnarray}
 {\cal M}_{\rm ps}(x) &=&
  \overline{q}_{1,c\alpha}(x) \gamma_{5,\alpha\beta} q_{2,c\beta}(x)
  \qquad \mbox{for pseudo scalar meson}, 
  \\
 {\cal M}_{{\rm v},k}(x) &=&
  \overline{q}_{1,c\alpha}(x) \gamma_{k,\alpha\beta} q_{2,c\beta}(x)
  \qquad \mbox{for vector meson}, 
\end{eqnarray}
with applying six set of hopping parameters;
first three sets for
$\kappa_{1}=\kappa_2$ ($\pi$ and $\rho$ mesons) 
with taking a number from 
$\{0.1678, 0.1665, 0.1640\}$, and 
another three sets for 
$\kappa_{1}>\kappa_2$ ($K$ and $K^\ast$ mesons) 
with taking different two numbers from 
$\{0.1678, 0.1665, 0.1640\}$. 
Assuming the following functional forms 
for the  pseudo scalar meson mass squared and
for the vector meson mass,
\begin{eqnarray}
 && (m_{ps}a)^2 = {B\over 2} \left(
			   {1\over \kappa_1} - {1\over \kappa_c} 
			  \right)
 + {B\over 2}\left(
	      {1\over \kappa_2} - {1\over \kappa_c} 
	     \right), \\
 && (m_{v}a) = C + {D\over 2} \left(
				{1\over \kappa_1} - {1\over \kappa_c} 
			       \right)
 + {D\over 2}\left(
	      {1\over \kappa_2} - {1\over \kappa_c} 
	     \right),
\end{eqnarray}
we obtained critical hopping parameter
$\kappa_c=0.1693$, and 
the physical parameter,
$\kappa_{phys}=0.1691$, from
$(m_\pi a/m_\rho a)=(135/770)$.
The lattice scale is determined so as to be $a=0.1420$ fm 
from the physical $\rho$ meson mass.
The parameter for the strange quark mass is determined as $\kappa_s=0.1643$
from the physical $K$ meson mass (494 MeV).

\section{Results and Discussion}

\begin{table}[b]
 \centering \leavevmode 
 \begin{tabular}{cccccccc}
  \hline
  $m_\pi$ & $m_\rho$ & $m_K$ & $m_{K^\ast}$ &
  $m_p$ & $m_{\Xi^0}$ & $m_\Lambda$ & $m_{\Sigma^+}$ \\
  \hline
  367(1) & 811(4) & 552.6(5) & 882(2) &
  1164(7) & 1379(6) & 1263(6) & 1312(6) \\
  \hline
  \end{tabular}
 \caption{Hadron masses, given in units of MeV,
 measured from the lattice QCD simulation.
 The number in the parenthesis shows the errorbar in the last digit.}
 \label{masses}
\end{table}
Table~\ref{masses} lists the hadron masses measured from 
the present lattice QCD simulation. 
1283 
gauge configurations are used to calculate the 
hadron masses and wave functions. 
 (17 
exceptional configurations are not used out of
  totally 1300 gauge configurations.)
We note that the present results for the baryon masses 
provide the correct order of particular threshold energy of 
two baryon states in the strangeness $S=-2$ sector; 
$E_{th}(\Lambda\Lambda)=2525(11)$MeV
(this channel is not allowed in the present case because of isospin
conservation), 
$E_{th}(N\Xi)=2544(12)$MeV,
$E_{th}(\Lambda\Sigma)=2575(11)$MeV, and 
$E_{th}(\Sigma\Sigma)=2624(11)$MeV. 
This 
warrants the desirable asymptotic behavior 
of the wave function.

\begin{figure}[t]
 \centering \leavevmode
 \includegraphics[width=.8\textwidth]{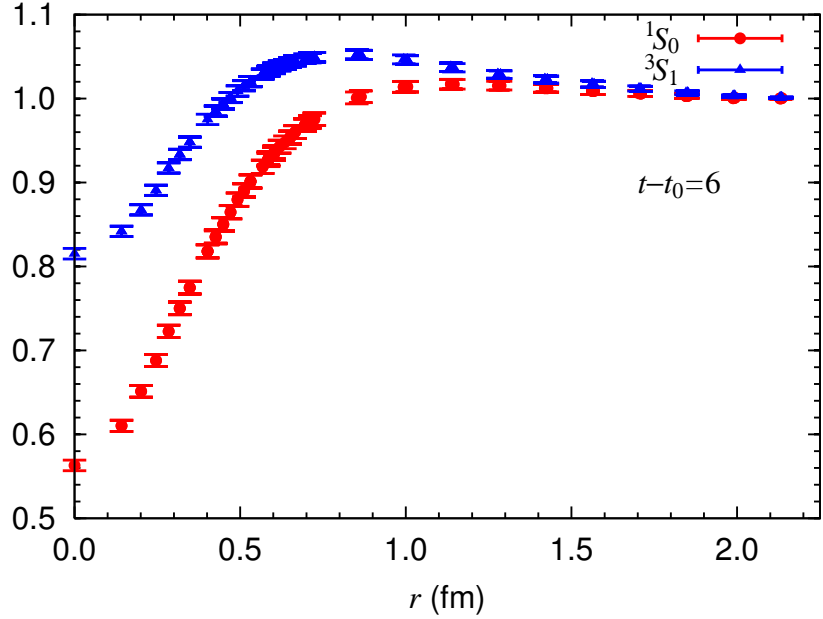}
 \caption{The radial wave function of $p\Xi^0$,
 in $^1S_0$ (circle) and $^3S_1$ (triangle) channels,
 obtained from the lattice QCD at $t-t_0=6$.
 }
 \label{wave}
\end{figure}
Figure~\ref{wave} shows the wave function obtained 
at the time slice $t-t_0=6$. 
The $^1S_0$ ($^3S_1$) channel is plotted by circles (triangles), 
which are normalized at the spatial boundary $\vec{r}=(32/2,0,0)$. 
All the data are taken into account for $r\lsim 0.7$ fm, while 
only the data on the $x$-, $y$-, and $z$-axis and their nearest neighbors 
are used to plot for the outer region. 
As seen in the Figure, the wave functions are suppressed in the 
short distance region, and a slight enhancement is found in the 
medium range region for both $^1S_0$ and $^3S_1$ channels. 
There is sizable difference between the $^1S_0$ and $^3S_1$,
particularly of suppression in the short distance, 
suggesting that the $^1S_0$ channel has stronger repulsive core. 

\begin{figure}
 \centering \leavevmode
 \includegraphics[width=.9\textwidth]{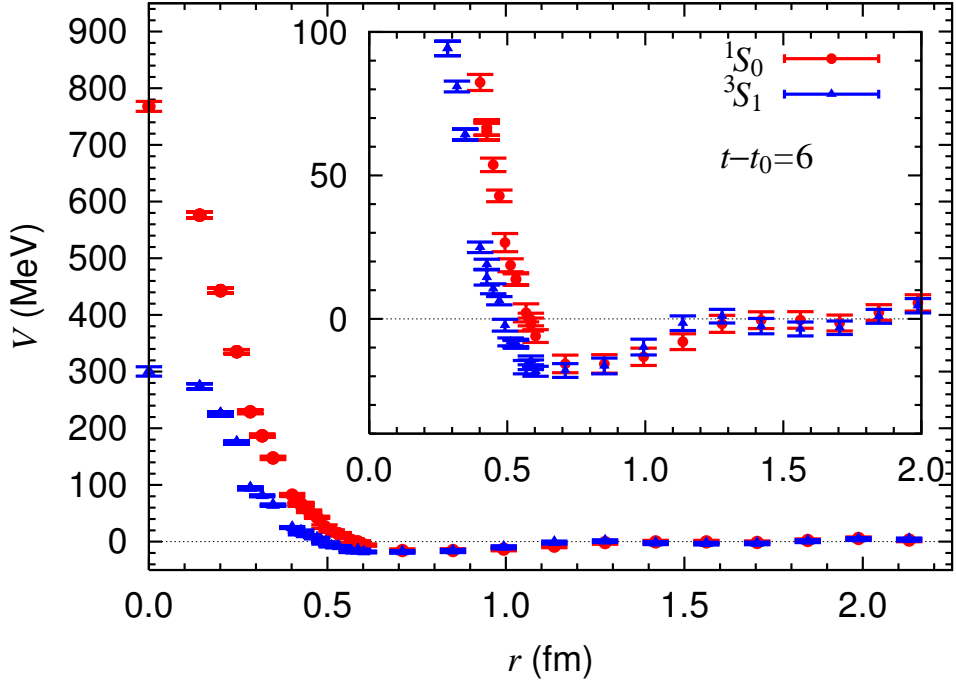}
 \caption{The effective central potential for $p\Xi^0$,
 in the $^1S_0$ (circle) and $^3S_1$ (triangle),
 obtained from the wave function at time slice $t-t_0=6$. 
 The inset shows its enlargement. 
 }
 \label{pot}
\end{figure}
Figure~\ref{pot} shows the (effective) central potentials 
for $p\Xi^0$ 
in the $^1S_0$ and $^3S_1$ channels. 
These results are still preliminary, since 
the potentials are obtained by assuming $E=0$:
 The energy should be determined by 
fitting the asymptotic behavior 
of the wave function with the use of 
the Green's function $G(\vec{r},E)$ 
which is a solution of the Helmholtz equation on 
the lattice\cite{Aoki,Luscher}. 
Preliminary calculation suggests that $E$ would be 
small 
negative values for both $^1S_0$ and $^3S_1$ channels, 
similar to the case of 
the $NN$ potential\cite{Ishii,IshiiProc}.

\begin{figure}[t]
 \centering \leavevmode
 \begin{minipage}[t]{0.49\textwidth}
  \includegraphics[width=.98\textwidth]{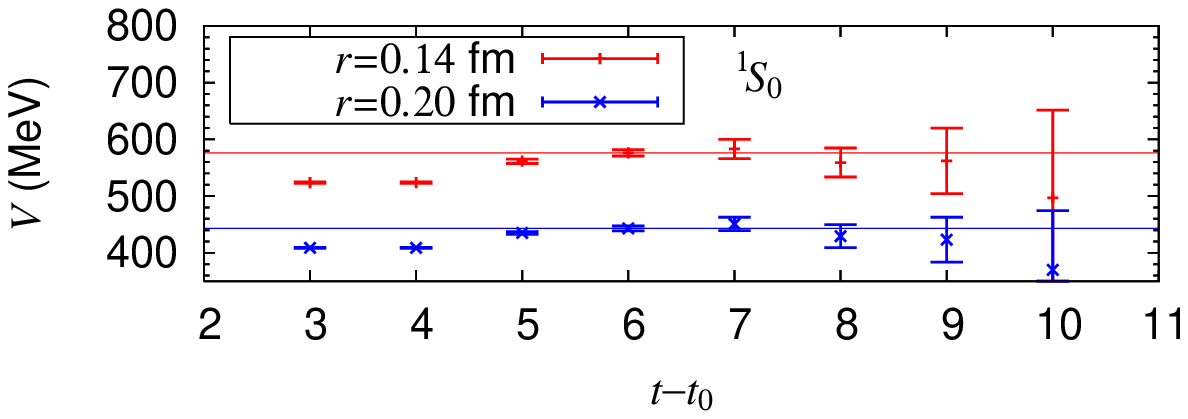}
  \includegraphics[width=.98\textwidth]{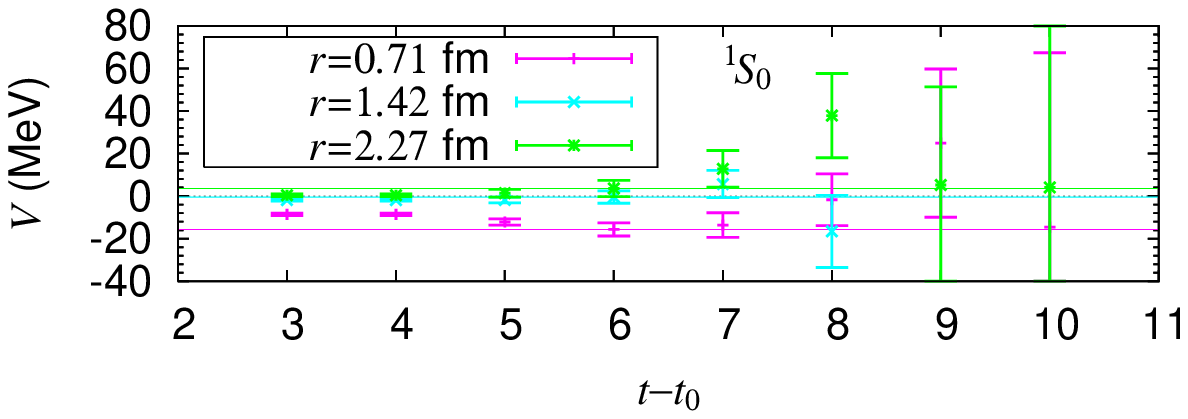}
  \footnotesize 
 \end{minipage}
 \hfill 
 \begin{minipage}[t]{0.49\textwidth}
  \includegraphics[width=.98\textwidth]{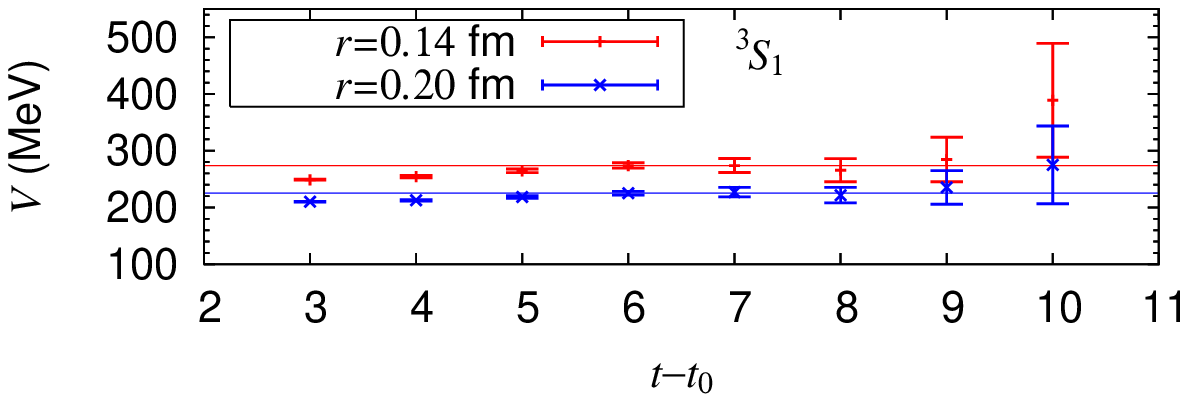}
  \includegraphics[width=.98\textwidth]{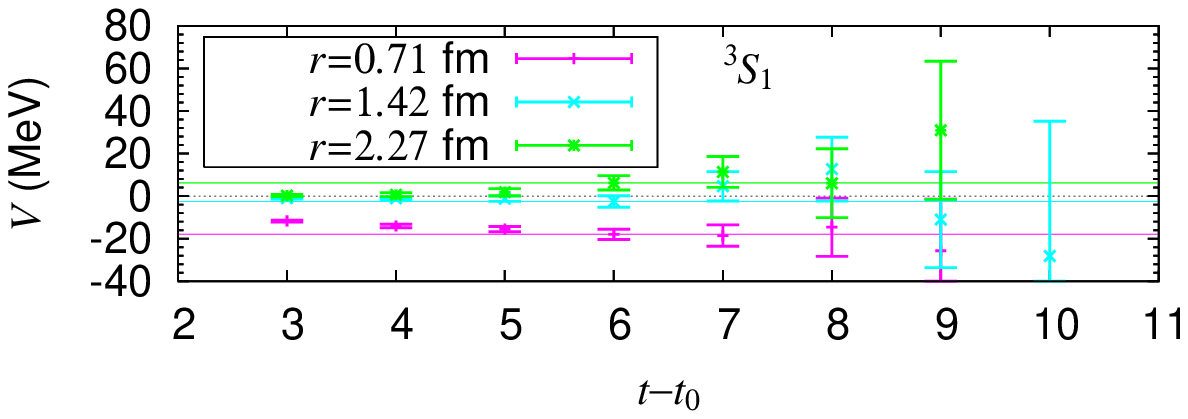}
 \end{minipage}
 \caption{The time-slice dependence of the potential 
 in the $^1S_0$ (left-hand-side) and $^3S_1$ (right-hand-side) channel 
 for several radial distance $r$. 
 }
 \label{pots}
\end{figure}

In order to see the ground state saturation of the present results, 
we plot, in Fig.~\ref{pots},
 the time-slice dependence of the potential in both of the 
 $^1S_0$ (left-hand-side) 
 and $^3S_1$ (right-hand-side) 
 channels
at several radial distances; 
$r=0.14$, $0.20$, $0.71$, $1.42$, and $2.27$ fm. 
We can see that the saturation is achieved for $t-t_0\ge 6$ within 
errors. 

The present work is a first step toward the $YN$ and $YY$ 
potentials from the lattice QCD simulation. 
Systematic studies of the various channels 
such as $\Lambda N$, $\Sigma N$, $\Lambda\Lambda$, and so on 
are all 
interesting and important 
because they are intimately related not only to the structure 
of hypernuclei but also to the 
internal structure of neutron stars. 
We will present such studies in the near future.

\acknowledgments

Lattice QCD Monte Carlo calculation has been done with IBM Blue Gene/L
computer at KEK. 
H.~N. is supported by the Special Postdoctoral Researchers Program 
at RIKEN. 
This research was partly supported by Grants-in-Aid for Young Scientists 
(B) (No. 17740174) from the Japan Society for Promotion of Science
(JSPS), 
and by the Ministry of Education, 
Science, Sports and Culture, Grant-in-Aid 
(Nos. 13135204, 15540251, 15540254, 18540253, 19540261).

\end{document}